\documentclass[twocolumn]{aastex62}

\usepackage{graphicx}
\usepackage{amsmath}
\usepackage{xspace}

\newcommand{\hMsol}{h^{-1}\,{\rm M_\odot}}

\newcommand{\hMpc}{h^{-1}\,{\rm Mpc}}

\newcommand{\kms}{{\rm km\, s^{-1}}}
\newcommand{\ie}{{i.e.~}}
\newcommand{\eg}{{e.g.~}}

\newcommand{\HI}{H\texttt{I}\xspace}

\newcommand{\Lya}{Ly$\alpha$\xspace}

\newcommand{\lt}{<}
\newcommand{\UR}{underdense region\xspace}  
\newcommand{\pMpc}{\mathrm{pMpc}}
\newcommand{\avgf}{\langle f (r) \rangle}
\newcommand{\ULAS}{ULAS J1120+0641\xspace}
\newcommand{\LOS}{line of sight\xspace}  
\newcommand{\hp}{h_\mathrm{s}}
\renewcommand{\wp}{w_\mathrm{s}}

\newcommand{\enrico}[1]{#1}

\graphicspath{{./}}

\submitjournal{The Astrophysical Journal}

\shorttitle{Probing Reionization with Transmission Spikes}
\shortauthors{Garaldi et al.}

\begin{document}

\title{Constraining the Tail-End of Reionization Using Lyman-$\alpha$ Transmission Spikes}

\correspondingauthor{Enrico Garaldi}
\email{egaraldi@mpa-garching.mpg.de}

\author[0000-0002-6021-7020]{Enrico Garaldi}
\affiliation{Max-Planck-Institut f\"ur Astrophysik, Karl-Schwarzschild-Str. 1, 85741 Garching, Germany}
\affiliation{Argelander Institut f\"ur Astronomie der Universit\"at Bonn, Auf dem H\"ugel 71, 53121 Bonn, Germany}

\author{Nickolay Y. Gnedin}
\affiliation{Particle Astrophysics Center, Fermi National Accelerator Laboratory, Batavia, IL 60510, USA}
\affiliation{Kavli Institute for Cosmological Physics, The University of Chicago, Chicago, IL 60637 USA}
\affiliation{Department of Astronomy \& Astrophysics, The University of Chicago, Chicago, IL
60637 USA}

\author{Piero Madau}
\affiliation{Department of Astronomy \& Astrophysics, University of California, 1156 High Street, Santa Cruz, CA 95064, USA}

\begin{abstract}
{We investigate Lyman-$\alpha$ transmission spikes at $z>5$ in synthetic quasar spectra, and discuss their connection to the properties of the intergalactic medium and their ability to constrain reionization models. We use state-of-the-art radiation-hydrodynamic simulations from the Cosmic Reionization On Computers series to predict the number of transmission spikes as a function of redshift, both in the ideal case of infinite spectral resolution and in a realistic observational setting. Transmission spikes are produced in highly-ionized underdense regions located in the vicinity of UV sources. We find that most of the predicted spikes are unresolved by current observations, and show that our mock spectra are consistent with observations of the quasar 
\ULAS in about $15$\% of the realizations. The spike height correlates with both the gas density and the ionized fraction, but the former link is erased when synthetic spectra are smoothed to realistically achievable spectral resolutions. There exists a linear relationship between spike width and the extent of the associated underdense region, with a slope that is redshift-dependent. In agreement with observations, the spike transmitted flux is suppressed
at small distance from bright galaxies as these reside in overdense regions. We argue that this anti-correlation can be used to constrain large-scale density modes.
}
\end{abstract}

\keywords{galaxies: formation -- galaxies: high-redshift -- intergalactic medium --  quasars: absorption lines -- cosmology: observations -- dark ages, reionization, first stars}

\section{Introduction}
\label{sec:intro}

The hydrogen stored in the intergalactic medium (IGM) transformed from completely neutral to a highly-ionized plasma at redshift $5.5 \lesssim z \lesssim 10$ \citep{Planck2018cosmo} in what is known as the Epoch of Reionization (EoR). It is widely believed that the ionizing photons that caused this transition were produced by early star-forming galaxies \citep[see \eg][]{madau99,gnedin00,haardt12}, with 
active galactic nuclei \citep{Haardt&Madau1996,Kulkarni+2018} and X-ray binaries \citep{Eide+2018} playing only a minor role at redshifts $z>5$
(but see \citealt{madau15}).

There are both empirical and theoretical arguments supporting this picture. Observationally, the Cosmic Microwave Background (CMB) anisotropies place the mid point of reionization at $7.64\pm 0.74$ \citep[if symmetric,][]{Planck2018cosmo}, while the evolution of the effective optical depth of the Lyman-$\alpha$ (hereafter \Lya) forest \citep[\eg][]{Fan+2006} and the sudden change in the density of detected \Lya-emitting systems \citep[\eg][]{Ota+2010,Pentericci+2011,Mason+2018} constrain the tail-end of reionization (where individual ionized bubbles overlap) to occur at redshift $5.5 \lesssim z \lesssim 6.5$. 
Additional constraints at various cosmic times have been obtained from modelling the near-zones of high-redshift quasars \citep[\eg][]{Schroeder+2013, Davies+2018} and from the dark pixel statistics in quasar absorption spectra \citep[\eg][]{McGreer+2011}.

 The amount of UV photons produced by galaxies, however, appears to be insufficient to fully ionize intergalactic 
 hydrogen if their escape fraction into the IGM is of order a few per cent, similar to the one typically observed in the local Universe \citep[see \eg][]{Marchi+2017}. Some lower-redshift galaxies exhibit escape fraction in excess of $10$\% \citep[and as large as $\sim 60$\%, ][]{Vanzella+2018}, but it remains unclear whether these are representative of the sources driving the EoR. Models typically assume larger escape fractions at higher redshift, although a consensus on the physical mechanism causing such evolution has not yet been reached. 

Historically, high-redshift quasars have been one of the most-powerful probes of the EoR, as they provide bright flashlights that illuminate the IGM along the line of sight to Earth. The most prominent and ubiquitous absorption line in quasar spectra is the \Lya transition of neutral hydrogen. The cross section for such process is very large, so that even a modest hydrogen neutral fraction of $x_\mathrm{\HI} \sim 10^{-4}$ produces complete absorption. For this reason, \Lya studies have been historically limited to the post-reionization universe and mainly to the range $3 \lesssim z \lesssim 4.5$, where the superposition of multiple individual absorption features due to isolated patches of not-fully-ionized hydrogen produce the so-called \Lya forest.

Recent improvements in observational facilities are boosting our ability to probe the tail-end of the reionization process using QSO spectra. In particular, better spectral resolution and sensitivity, coupled with the discovery of QSOs at higher and higher redshift\footnote{There are now more than 150 detected QSOs at $z>6.0$ according to \citet{QSO_website} and the current record-holder is the QSO at $z = 7.54$ discovered by \citet{Banados+2018}.}, are unveiling features at the edge of the reionization period.
\citet{Barnett+2017} presented a high-resolution spectrum of the QSO \ULAS at $z_\mathrm{QSO} = 7.084$ obtained with a 30h integration using the X-shooter instrument mounted on the Very Large Telescope. The only detected transmission blueward of the rest-frame \Lya wavelength (with signal-to-noise ratio larger than $5$) comes from $7$ narrow spikes in the \Lya forest. These span the redshift range $5.858 < z <6.122$ and are located at the low-redshift edge of a long Gunn-Peterson trough (of size $\sim 240 \, \hMpc$) extending all the way to the QSO proximity zone at $z_\mathrm{pz} = 7.04$.

In a first attempt to extract information from \enrico{similar observations, \citet{Gallerani+2006, Gallerani+2008} used a semi-analytical model to relate the transmission windows and absorption gaps to the IGM ionized fraction, suggesting that transmission region are spatially associated with galaxies. If this result is confirmed, the radiation field should be enhanced in the regions where spikes are produced. We anticipate here, however, that such geometrical argument does not directly entail a boosted transmission around galaxies since the sources of ionizing photons reside in overdense regions, where recombination is more efficient at suppressing the transmitted flux.}

\enrico{More recently, \citet{Chardin+2018} combined the spectrum of \ULAS with a set of hydrodynamical simulations post-processed with a radiative transfer code, showing} that the number of spikes can, in principle, constrain the timing of reionization. Their work, however, relies on simulations of modest box sizes
that neglect any coupling between gas dynamics and radiation, an effect that may be important for a detailed treatment of \Lya forest features. 

In order to fully exploit available and forthcoming observations, in this Paper we complement their analysis using three radiation-hydrodynamical simulations from the Cosmic Reionization On Computers suite. We investigate the physical conditions and processes producing transmission spikes during the EoR and present the first detailed theoretical prediction for the correlation between spikes and neighboring galaxies. The text is organized as follows. After describing the simulations employed and the characterization procedure for the synthetic spectra (\S~\ref{sec:methods}), we present our main findings in \S~\ref{sec:results}, where we compare the simulations with a set of recent observations. Finally, we provide a summary and a final discussion of the implications of our results in \S~\ref{sec:conclusions}.

\section{Methods}
\label{sec:methods}
Numerical studies of the EoR face formidable difficulties. The necessity to include a proper treatment of radiation transport (RT) increases the dimensionality of the problem and forces a compromise between simulation volume and resolution. But even after the ideal balance has been found, the comparison of simulated and real data is not straightforward since the simulated physical quantities are not directly observable. Hence, simulations have to be post-processed in order to provide a meaningful comparison with observations. In this section we describe the simulation suite employed in this work as well as the production and subsequent characterization of synthetic absorption spectra.

\subsection{CROC simulations}
\label{sec:croc}

In this work we use three simulations from the Cosmic Reionization On Computers (CROC) suite of \citet{Gnedin+2014}.
CROC is a set of radiation-hydrodynamical simulations performed using the Adaptive Refinement Tree (ART) code 
\citep{Kravtsov+1997}. In particular, we employ simulations from the Caiman series, which include numerical convergence corrections \citep[see][]{Gnedin2016} and a better time sampling of the tail-end of reionization ($5 \lesssim z \lesssim 6.5$). We refer the interested reader to the CROC design paper \citep{Gnedin+2014} and summarize in the following only the main features of the simulation set. 

\begin{itemize}
\item The heating and cooling rates of hydrogen and helium are computed self-consistently during the simulation run,
without assuming photoionization or collisional equilibrium. In particular, the approximation of \citet{Gnedin&Hollon2012} is used to determine the metallicity dependence of the heating and cooling functions as a function of the local radiation field. 

\item Molecular hydrogen production and destruction is implemented using a fitting function calibrated against high-resolution self-consistent simulations \citep{Gnedin&Draine2014}. 
\item Star formation is included through an empirical sub-grid model, following a linear Kennicut-Schmidt relation \citep{Schmidt1959,Kennicutt1998} and assuming a typical star-formation timescale of $\tau_\mathrm{SF} = 1.5 \, \mathrm{Gyr}$. The feedback from stars is implemented following the standard `delayed cooling' model \citep{Stinson+2006}, 
while the ionizing radiation produced during their life is computed combining a \citet{Kroupa2001} initial mass function with the Starburst99 model \citep{Leitherer+1999} for the spectral shape.

\item The CROC simulations are run employing the Optically Thin Variable Eddington Tensor (OTVET) approximation \citep[][updated to suppress numerical diffusion]{Gnedin&Abel2001} for the radiation transport of UV (including ionizing) photons emitted by stars. Other sources of ionizing radiation (such as quasars, bremsstrahlung and helium recombination radiation) are included as a background either because they are weakly clustered or too rare (at the redshift covered by the simulations) to significantly influence a randomly-selected region of the universe. The distortion of the background spectrum associated with the opacity of the optically thick Lyman-limit systems is computed employing the fit by \citet{Songaila&Cowie2010}.

\item The free parameters in the physical models have been calibrated against the observed galaxy UV luminosity function in the redshift range $5 \lesssim z \lesssim 10$ (that constrains the two parameters involved in star-formation), and against the Gunn-Peterson optical depth of the \Lya forest in the spectrum of high-$z$ quasars (that constrains the ionizing photons escape fraction at the simulation resolution).

\item Density fluctuations on scales larger than the box size are treated using the `DC mode' formalism presented in \citet{Gnedin+2011}.

\end{itemize}

\begin{figure}
\includegraphics[width=\columnwidth]{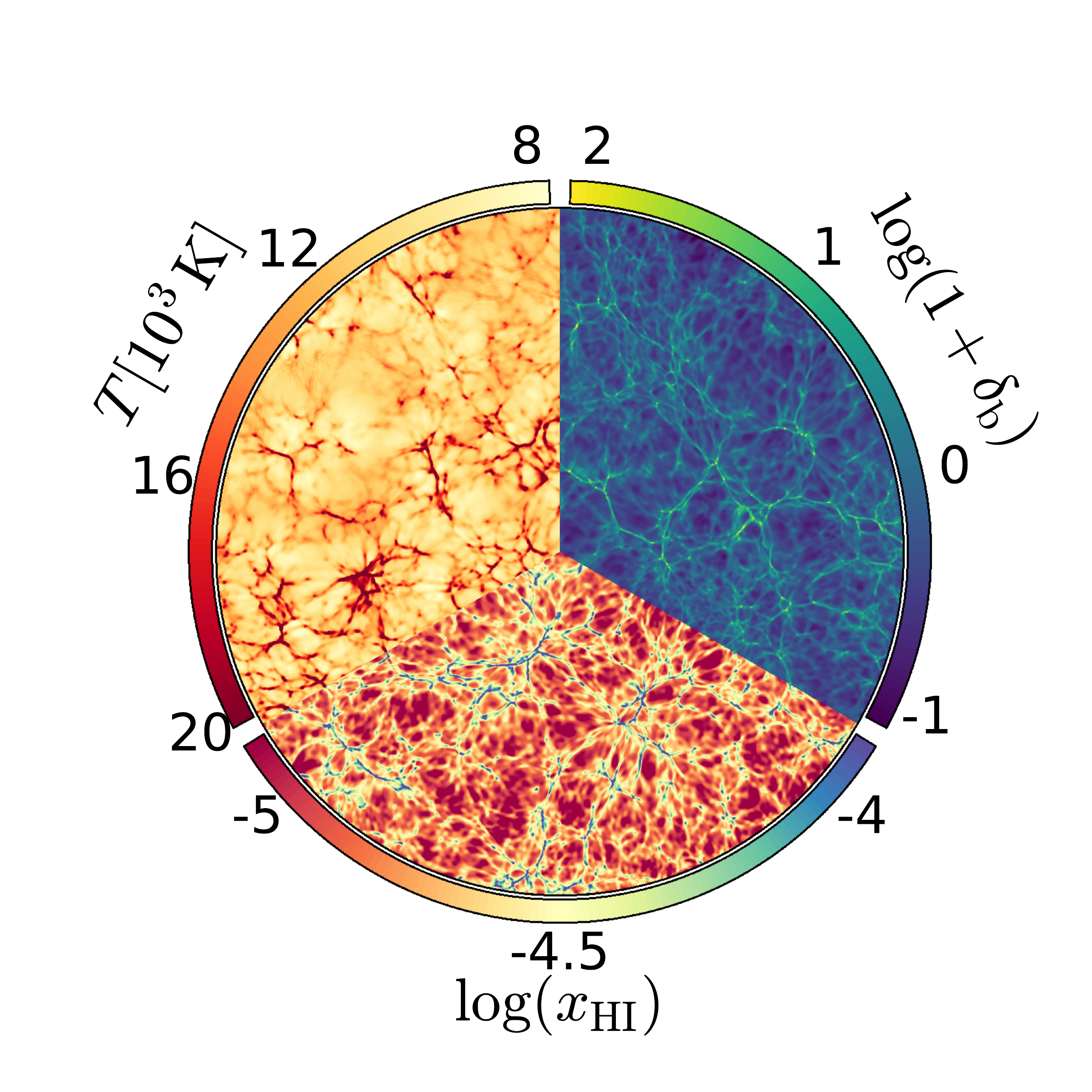}
\caption{Slice through the simulation box. Subregions show the gas overdensity $1+\delta_\mathrm{b}$, \HI fraction $x_\mathrm{\HI}$, and temperature $T$.}
\label{fig:pie}
\end{figure}

In this work we employ three numerical simulations with box size $L_\mathrm{box} = 80 \, \hMpc$ (comoving). The computational box is initially discretized in a Cartesian grid containing $2048^3$ elements, and is subsequently adaptively refined to maintain spatial resolution approximately constant at $100\,\mathrm{pc}$ in physical units. At redshift 
$z=6$, the corresponding Nyquist frequency $k_{\rm Ny}$ (in velocity units) is $\approx 0.27 \, \mathrm{s \, km^{-1}}$ ($\pi/k_\mathrm{Ny} \approx 12\,\mathrm{km\, s^{-1}}$), 
which is sufficient to model the IGM features we are interested in. The simulations assume a present day Hubble constant $H_0 = 68.14 \, \mathrm{km} \, \mathrm{s}^{-1} \, \mathrm{Mpc}^{-1}$, while the density parameters for baryonic matter, total matter and cosmological constant are $\Omega_\mathrm{m} = 0.3036$, $\Omega_\mathrm{b} = 0.0479$ and $\Omega_\Lambda = 0.6964$, respectively \citep[other cosmological parameters and simulation details are given in][]{Gnedin+2014}.
 
For the current analysis, we employ 13 redshift bins spanning the redshift range $5.0 \lesssim z \lesssim 6.5$. In each of these bins simulation data were remapped onto a uniform, $1024^3$ (cell size $= \pi/k_{\rm Ny}$) grid for the purpose of making synthetic absorption spectra.
The three simulations we analize follow the evolution of the same initial conditions but differ in their values of the so-called `DC mode', the density fluctuation $\delta_\mathrm{box}$ on the scale of the simulation box \citep[which is not zero, since the box is finite;][]{Pen+1997,Sirko+2005,Gnedin+2011}. Specifically, we impose $\delta_\mathrm{box}$ to be -1, 0, and +1 times the theoretically expected rms density fluctuation in a cubic region of $80\,\hMpc$ on a side (i.e. $\delta_\mathrm{box} = -\sigma_\mathrm{box}$, 0, and $+\sigma_\mathrm{box}$ respectively). This allow us to explore the effect of large-scale density fluctuations in a fully-consistent way.

A visual impression of the simulations employed is given in Figure \ref{fig:pie}, where we show three different quantities across a slice through one simulation box: the baryonic overdensity $\delta_\mathrm{b} \equiv \rho_{b} / \bar{\rho}_{b} - 1$ (with $\rho_{b}$ being the baryonic density field and $\bar{\rho}_{b}$ its value averaged over the entire simulation), the \HI fraction $x_\mathrm{\HI}$, and the gas temperature $T$.

\subsection{Synthetic spectra}
\label{sec:spectra}

\begin{figure}
\includegraphics[width=\columnwidth]{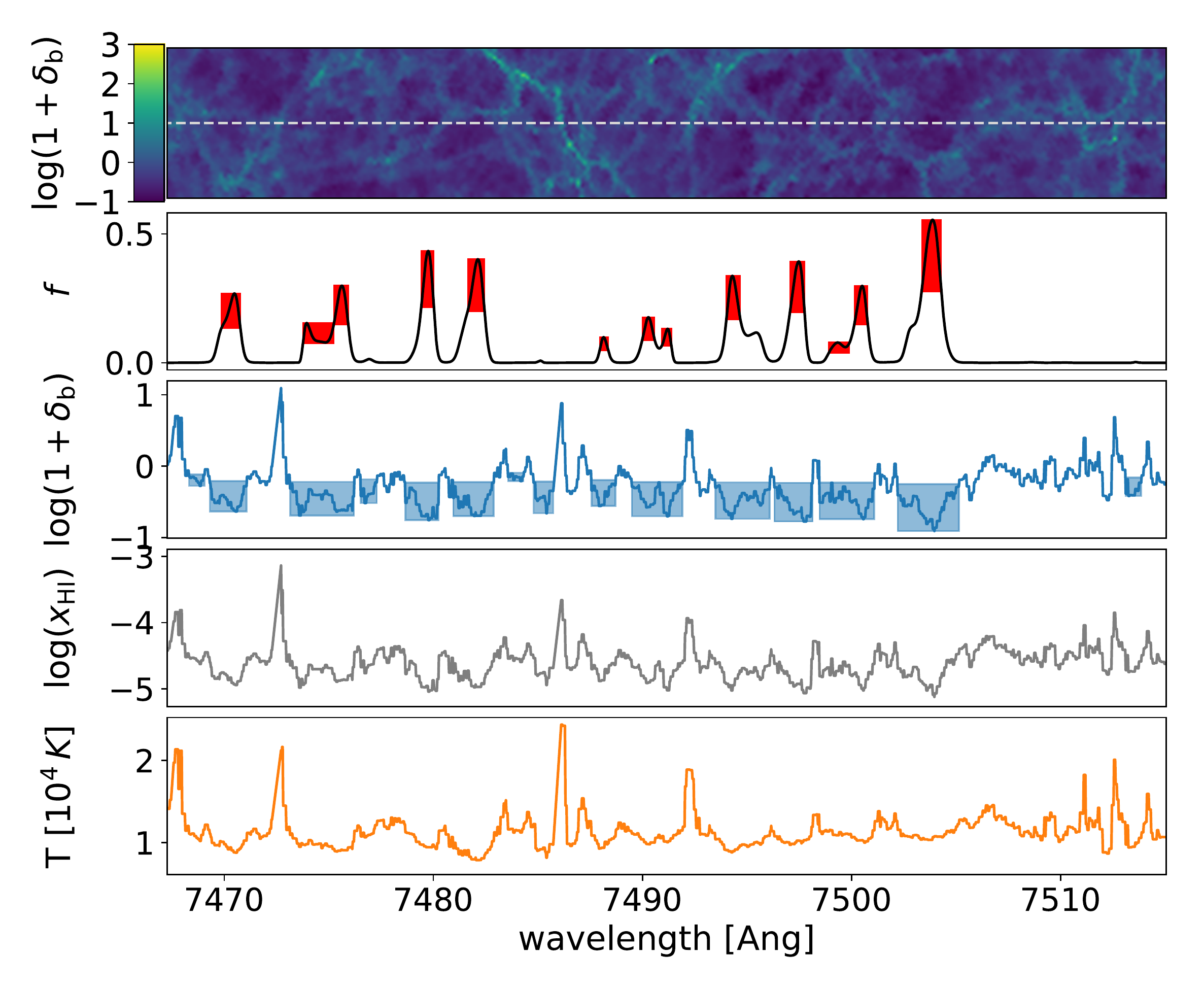}
\caption{A random line of sight at $z = 5.6$ extracted from our simulations. The top panel shows the baryonic overdensity in a slice through the simulation box centered on the selected \LOS (dashed line). The remaining four panels show, from top to bottom, the normalized transmitted flux $f$, the baryonic overdensity $1+\delta_\mathrm{b}$, the hydrogen neutral fraction $x_\mathrm{\HI}$, and the gas temperature $T$ along the \LOS. The wavelength 
is given in the observer rest-frame. Red (blue) boxes highlight the identified spikes (underdense regions).}
\label{fig:showcase}
\end{figure}

We produce synthetic \HI \Lya absorption spectra by extracting the distribution of neutral hydrogen, gas temperature and velocity along random lines of sight  
of length $L_\mathrm{s} = 80 \, \hMpc$ within the simulation box. 
The resolution elements of such spectra are $1 \, \kms$, to ensure that all simulated features are heavily over-sampled and, hence, fully captured. Optical depths are transformed into (normalized) transmitted fluxes $f$ using a Voigt profile and including peculiar velocities, Doppler and thermal broadening. An example of a synthetic spectrum at redshift $z=5.6$ is shown in Figure
\ref{fig:showcase} (second row from the top). The baryonic overdensity surrounding the selected \LOS 
(dashed line) is reported in the top panel in a slice of size $24 \, \hMpc \times 1.8 \, \hMpc$ of the simulation box. The remaining panels show the physical properties of the gas along the \LOS, namely (from middle to bottom) gas overdensity, \HI fraction, and gas temperature. 

At redshift $z \lesssim 4.5$ a substantial fraction of the IGM is transparent to \Lya photons, giving origin to the \Lya forest. At earlier times, however, most of the IGM is opaque to 
these photons, producing complete absorption everywhere but in a handful of highly-ionized regions. The resulting spectrum shows only few transmission spikes (see Figure  \ref{fig:showcase}). 
Quantitatively, we identify the latter as local maxima in the transmitted flux. The spike height ($\hp$) is defined as the maximum (normalized) transmitted flux while the width ($\wp$) 
corresponds to the simply-connected set of pixels that have $f \geq \alpha \hp$, where $\alpha$ is an adjustable parameter set to $\alpha = 0.5$ \citep{Gnedin+2017}.

We associate each transmission spike to the physical properties of the co-spatial IGM. 
In order to link a single value of each gas property to a spike, we employ as a representative value the average over the spike width weighted by the transmitted flux in each pixel, \ie 
\begin{equation}
\bar{q} \equiv \frac{ \int f(\lambda) \, q(\lambda) \, \mathrm{d} \lambda \, }{ \int f(\lambda) \, \mathrm{d} \lambda } ~,
\end{equation}
where $q$ is a generic gas property and $\lambda$ is the wavelength. More specifically, in this work we associate to each spike the gas temperature $T$, baryonic overdensity $\delta_{\rm b}$, and neutral fraction $x_{\rm HI}$. 

\begin{figure}
\includegraphics[width=\columnwidth]{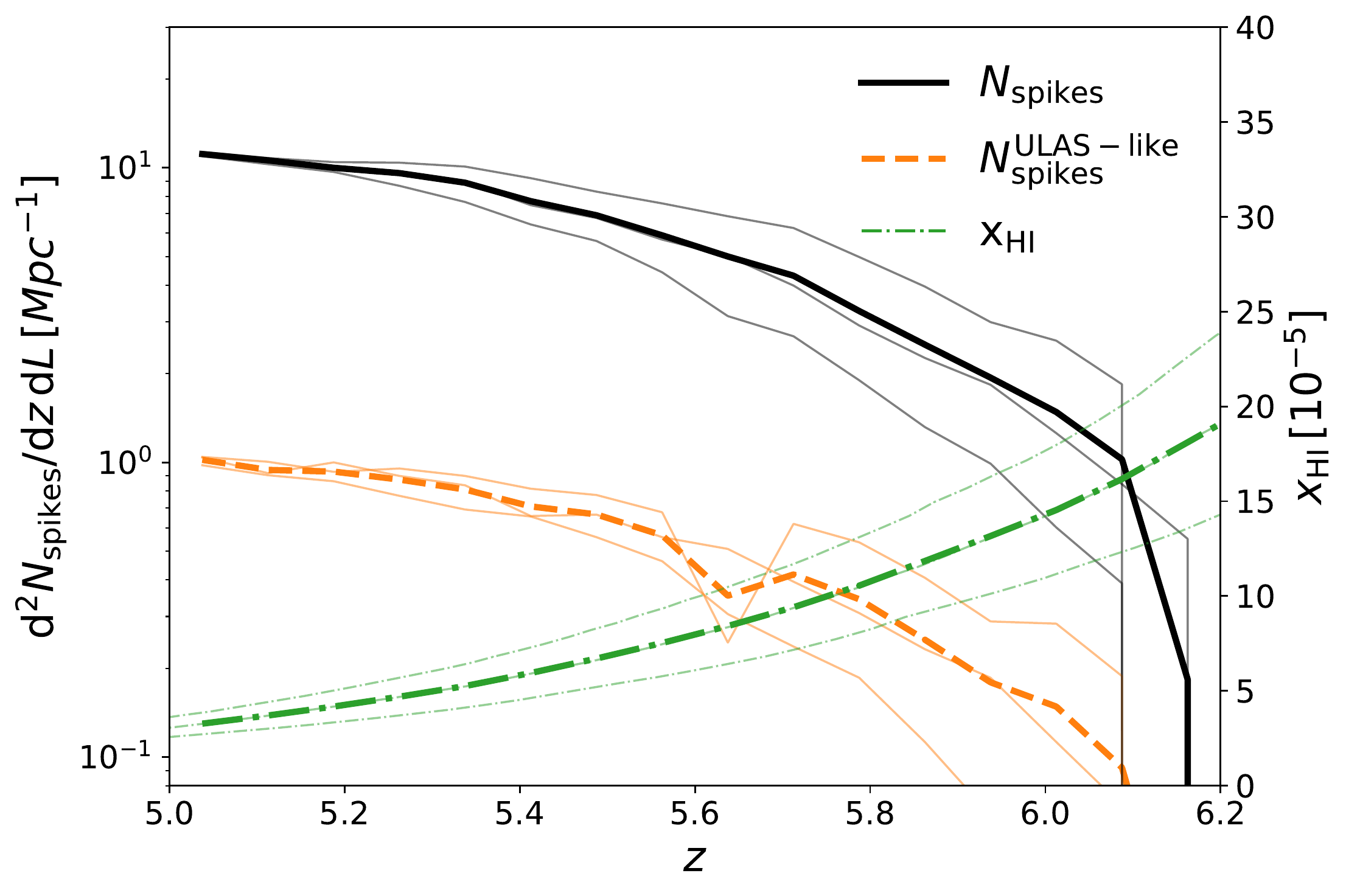}
\caption{Number of spikes per unit redshift per comoving $\mathrm{Mpc}$ as a function of redshift in an ideal case (\ie infinite spectral resolution and no noise, solid lines) and in 
a \ULAS-like observational settings (dashed lines). We also show the volume-averaged neutral fraction for the three simulations in our suite (dot-dashed lines). Thin lines correspond 
to results obtained from individual boxes, while the thick ones indicate the average over all boxes.
\label{fig:evolution}}
\end{figure}

\begin{figure*}
\includegraphics[width=\textwidth]{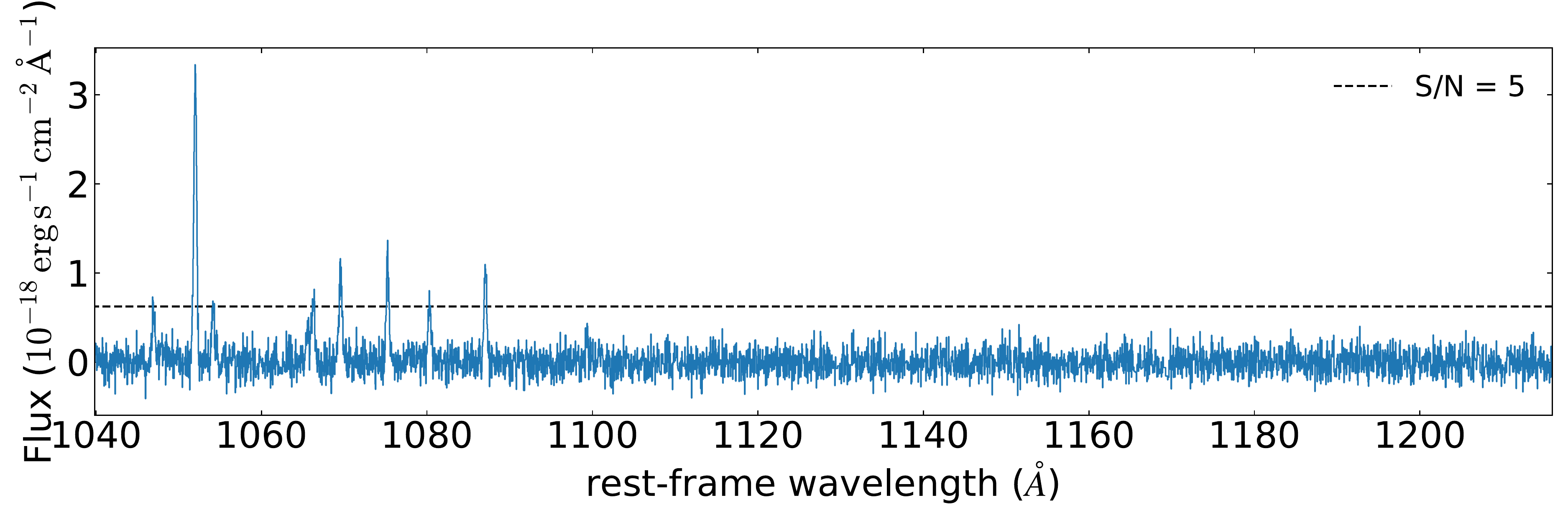}
\caption{Synthetic spectrum extracted from our simulation by concatenating randomly-selected lines of sight extracted from the simulation output at the 
appropriate redshift. The dashed line marks the flux level $5\sigma$ above the noise. 
}
\label{fig:multibox_spectrum}
\end{figure*}

Similarly, we identify underdense regions 
along the \LOS. Starting from local minima in the overdensity ($\delta_{\rm min}$), we define the \UR as the (simply-connected) region having $\delta < \beta \delta_{\rm min}$, where $\beta$ is again an adjustable parameter. 
This provides us with two separate sets of \LOS portions that need to be matched. We do so by associating a spike to an \UR whenever the pixel with the maximum transmitted flux is within the \UR. 

The second (third) panel from the top of Figure \ref{fig:showcase} show the results of these procedures. The shaded boxes highlight the spikes (underdense regions) identified in the 
spectrum. Each box spans vertically the region $[1, \alpha]$ ($[1, \beta]$) times the maximum (minimum) value and covers horizontally the full spike (\UR) width. It appears 
clear that many underdense regions are not associated with any spike, while every spike is associated with an \UR. This qualitative result (based on a single sight-line) is investigated in more 
depth in the next section.

\begin{figure*}
\includegraphics[width=\textwidth]{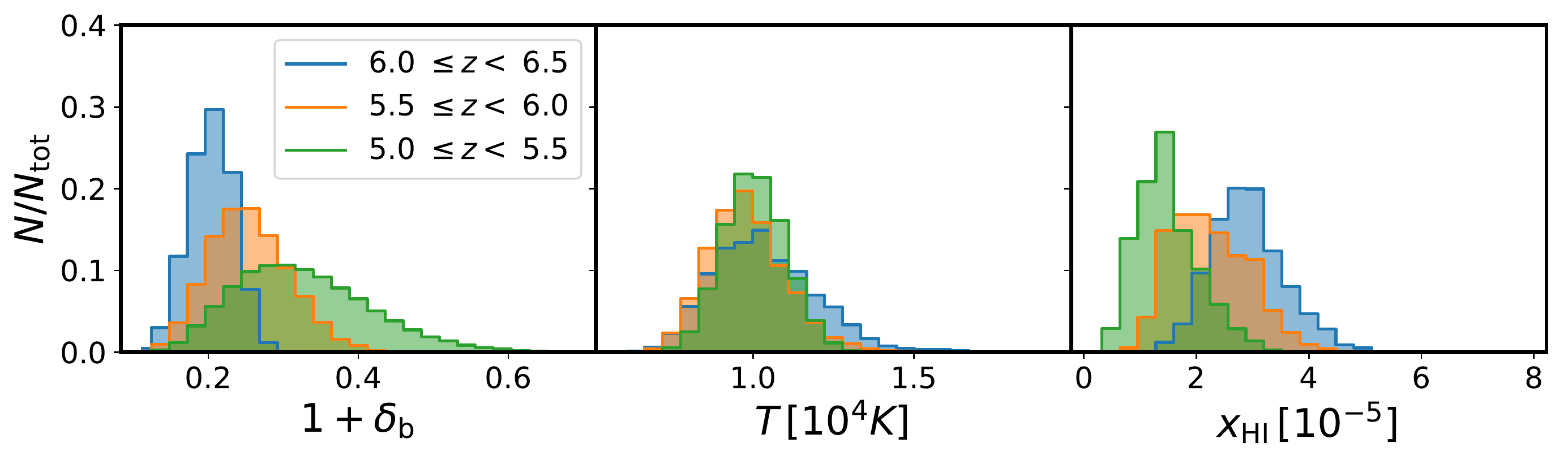}
\caption{Distribution of gas properties (baryonic overdensity $\delta_{\rm b}$, temperature $T$, and neutral fraction $x_{\rm HI}$, from left to right) associated 
with transmission spikes identified in simulated spectra within three redshift bins.}
\label{fig:peaks1D}
\end{figure*}

\begin{figure}
\includegraphics[width=\columnwidth]{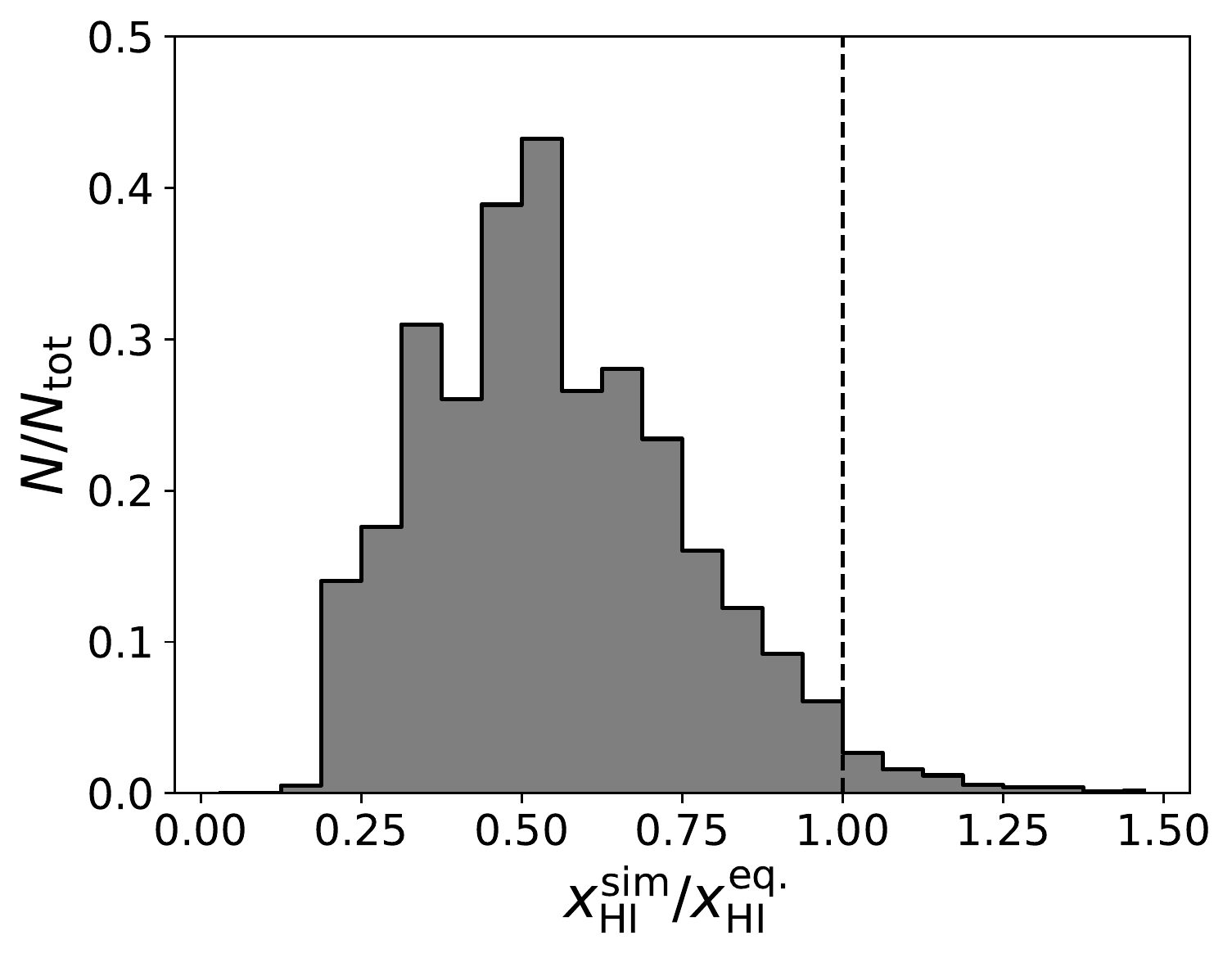}
\caption{{Distribution of ratios between the simulated gas neutral fraction associated with a transmission spike ($x_\mathrm{HI}^\mathrm{sim}$) and the value expected 
from ionization equilibrium with the average background radiation in a region of equal density ($x_\mathrm{HI}^\mathrm{eq.}$). The dashed line marks 
the position where the two values are equal. All redshift bins follow the same distribution and, hence, are shown together.}}
\label{fig:xHI_ratio}
\end{figure}

\begin{figure*}
\includegraphics[width=\textwidth]{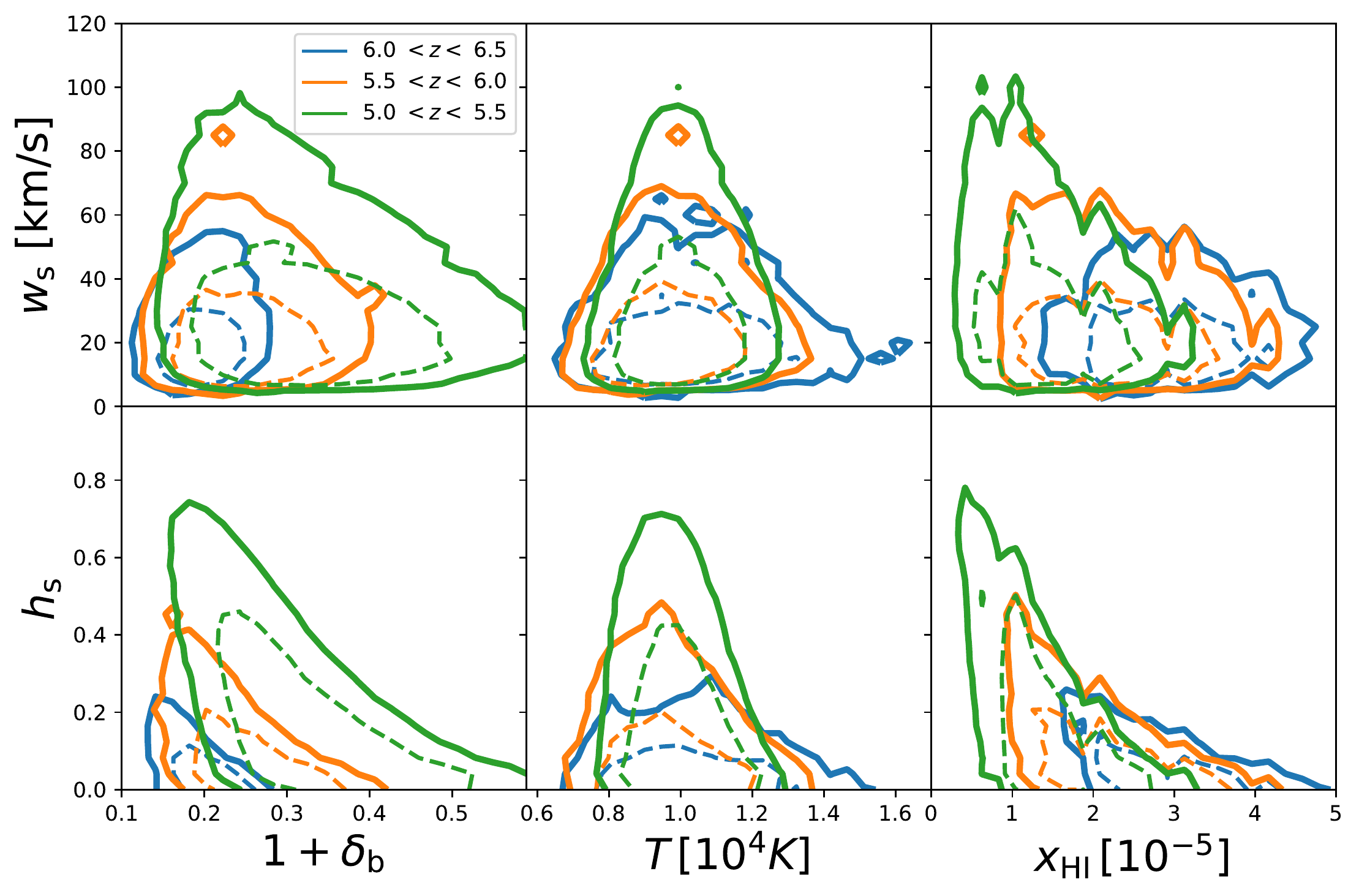}
\caption{Joint distributions of spike shape (characterized by their height $\wp$ in the top row, and their width $\hp$ in the bottom one) and associated gas properties: gas overdensity (left column), temperature (middle column) and HI fraction (right column). The simulated spectra have been analyzed in three different redshift bins. The solid (dashed) lines enclose the central 95 (68) per cent of the data.
}
\label{fig:peaks2D}
\end{figure*}

\section{Results}
\label{sec:results}

In this section we present the main findings of this work. We start by a careful physical characterization of the IGM underlying high-$z$ transmission spikes. Then, we move to study the average flux around bright sources, providing the first detailed predictions of such observable. 

\subsection{Spike number evolution}
\label{sec:number_evol}
We start our analysis of the simulated spectra by characterizing the occurrence of transmission spikes as a function of redshift. They are identified as described in \S~\ref{sec:spectra} and the resulting number $N_\mathrm{spikes}$ per unit redshift per comoving $\mathrm{Mpc}$ is shown in Figure \ref{fig:evolution} (solid lines)
as a function of redshift. This number increases approximately linearly with redshift, driven by the decreasing average neutral fraction. 
We will investigate in more details this correlation in \S~\ref{sec:igm-prop}.

Instrumental and observational effects, like finite spectral resolution and noise, reduce the measured value of $N_\mathrm{spikes}(z)$. As an example, we simulate 
observations with the same characteristics as those of the quasar \ULAS \citep{Barnett+2017}. We do so by smoothing our synthetic spectra with a boxcar filter 
that mimics an instrument of finite spectral resolution $R = c/ \Delta v  \approx 10000$, and adding Gaussian
white noise with rms $\sigma = 0.125 \times 10^{-18} \, \rm erg \, s^{-1} \, cm^{-2} \, \AA^{-1}$. The corresponding number of detected (\ie with signal-to-noise larger 
than 5) spikes $N_\mathrm{spikes}^\mathrm{ULAS-like}(z)$ is reported in Figure\ \ref{fig:evolution} (dashed lines).
Including these effects reduces the detected number of spikes by approximately an order of magnitude across all redshifts investigated. In the next section, 
we will use these `ULAS--like' spectra to determine in a quantitative way if our simulations are in agreement with spikes observed in the spectrum of \ULAS.

\subsection{Comparison with \ULAS}
Before delving into the properties of high-$z$ transmission spikes, we assess weather our simulations are able to produce synthetic spectra similar to the spectrum of \ULAS. In order to do so, 
we produce spectra of length $320 \, \hMpc$ (comoving), approximately corresponding to the distance between the quasar proximity zone and the onset of the Ly$\beta$ region. We do so by dividing 
the full $320 \, \hMpc$ pathlength into 8 segments (the number of simulation snapshots we have in this redshift interval) and concatenating randomly-selected $40 \, \hMpc$-long segments  
from consecutive simulation outputs at the appropriate redshift, assuming the quasar is located at $z_\mathrm{QSO} = 7.084$.
We thus account for light-cone effects albeit only in a piece-wise constant approximation. 
\enrico{We repeat this procedure 1000 times for each simulation box, for a total of 3000 spectra.} 
In the redshift covered by the spectrum ($5.86 < z < 7.04$), only few 
local features (transmission spikes) are present, and therefore the piece-wise constant approximation is not expected to introduce any artifact. 

We show in Figure \ref{fig:multibox_spectrum} one of these post-processed synthetic spectra that is compatible with \ULAS, together with 
the flux level $5\sigma$ above the noise. The latter is used as a threshold to identify spikes (\ie spikes are required to have signal-to-noise ratio larger than 5).
We consider our synthetic spectra compatible with that of \ULAS if they present $N_\mathrm{spikes} = 7 \pm 3$ spikes in the first (\ie lowest-redshift) $80 \, \hMpc$ 
and $N_\mathrm{spikes} \leq 1$ in the remaining $240 \, \hMpc$. The uncertainties on these numbers are obtained as the Poisson error (rounded to the nearest integer) 
of the observed number of spikes in \ULAS for each of the aforementioned redshift ranges. We find that $\approx 15\%$ of our synthetic spectra are compatible with 
the one observed by \citet{Barnett+2017}. This number, however, varies significantly between simulations using a different DC mode, \enrico{ranging from approximately $11$\% for the box with a positive DC mode, to $17$\% for the one with negative DC mode}. 

\subsection{Transmission spikes and IGM properties}
\label{sec:igm-prop}
This paper is mainly devoted to the theoretical investigation of the connection between transmission region in the high-$z$ \Lya forest and the associated IGM. 
Therefore, we do not include noise and instrumental effects in the following analysis in order to provide a general view on this topic, which can be 
adjusted to both current and future observations. Additionally, the latter are still too sparse to allow a proper comparison. 
Nevertheless, we discuss in Appendix \ref{sec:AppendixA} how the results presented in this section  are modified by the inclusion of observational effects.

In order to quantitatively study the physical connection between transmission spikes and the high-$z$ IGM, we start by investigating the 1D distribution 
of gas properties associated to transmission spikes in our synthetic spectra. This is shown in Figure\ \ref{fig:peaks1D} for three different 
redshift bins equally spaced  in the range $5.0 \leq z \lt 6.5$. All the spikes originate from regions that are underdense 
(\ie $1+\delta_\mathrm{b} < 1$) at all the times investigated, and become progressively emptier at higher redshifts. 
However, a low density IGM is not sufficient to produce a spike, since the gas needs to be also highly 
ionized.
The typical neutral fraction associated with spikes decreases from $x_{\rm HI} \approx 3 \times 10^{-5}$ 
at $6.0 < z < 6.5$ to $x_{\rm HI} \approx 1 \times 10^{-5}$ at $5.0 < z < 5.5$. At the redshift investigated, such combination of IGM properties is unlikely, 
explaining the rarity of the observed spikes. In particular, the production of these features requires underdense regions 
that experience a larger-than-average radiation field.
To substantiate this statement, we compute for each spike the ratio between the associated neutral fraction ($x_\mathrm{HI}^\mathrm{sim}$) and the value expected from 
ionization equilibrium with the average background radiation at the density of the spike. The latter amount to $x_\mathrm{HI}^\mathrm{eq.} = \bar{x}_\mathrm{HI} ( 1 + \delta_\mathrm{b})$, 
where $\bar{x}_\mathrm{HI}$ is the average neutral fraction in the simulation box. The distribution of such ratios is shown in Figure \ref{fig:xHI_ratio} as a single 
histogram, since we checked that the three redshift bins give consistent results. The vast majority (more than 98\%) of the spikes are produced in regions significantly 
more ionized than expected from ionization equilibrium with the average radiation field, \ie $x_\mathrm{HI}^\mathrm{sim} / x_\mathrm{HI}^\mathrm{eq.} < 1$.
We interpret this as a consequence of spikes production loci being preferentially 
close to a bright source of radiation. \enrico{This is consistent with the conclusions from \citet{Gallerani+2008}, based on a semi-analytical model, but} we will investigate this interpretation in more details in \S~\ref{sec:spikes-galaxy}.

Further information about the IGM and the sources of ionizing photons can be extracted from the shape of the spikes \citep{Garaldi+2019}.  \citet{Gnedin+2017} showed that the CROC simulations produce a realistic distribution of spike heights and 
widths, with the exception of the $\wp \gtrsim 800 \, \kms$ tail of the spike width distribution. We are therefore confident that the simulations 
employed in this work are suitable for our purposes.
In Figure \ref{fig:peaks2D} we plot the joint distributions of spike widths and heights with gas overdensity, temperature, and \HI fraction.  
The distributions show that the variables investigated are mostly uncorrelated, therefore making it hard to translate the shape of a spike into direct information about the underlying gas. Nevertheless, the very observation of a spike at a given redshift can be used constrain the underlying IGM physical properties. While these can cover a broad range at $5.0 < z < 5.5$, at even earlier epochs our results indicate that only a relatively narrow range of IGM overdensities  underlies the observed spikes. 

The spike height appears to be broadly correlated with the gas density, especially at lower redshifts, and to the underlying neutral fraction. In 
particular, $\hp$ is more sensitive to $x_{\rm HI}$ in the redshift range $5.5 < z < 6.0$: in this interval the contours are more tilted with respect to the axes, while in the other two redshift ranges the contours are almost parallel to either the horizontal (at $6.0 < z < 6.5$) or the vertical (at $5.0 < z < 5.5$) axis. The distribution of IGM temperatures is largely insensitive to spike width and height, as well as redshift.

\subsection{Spike width and underdense regions}

\begin{figure}
\includegraphics[width=\columnwidth]{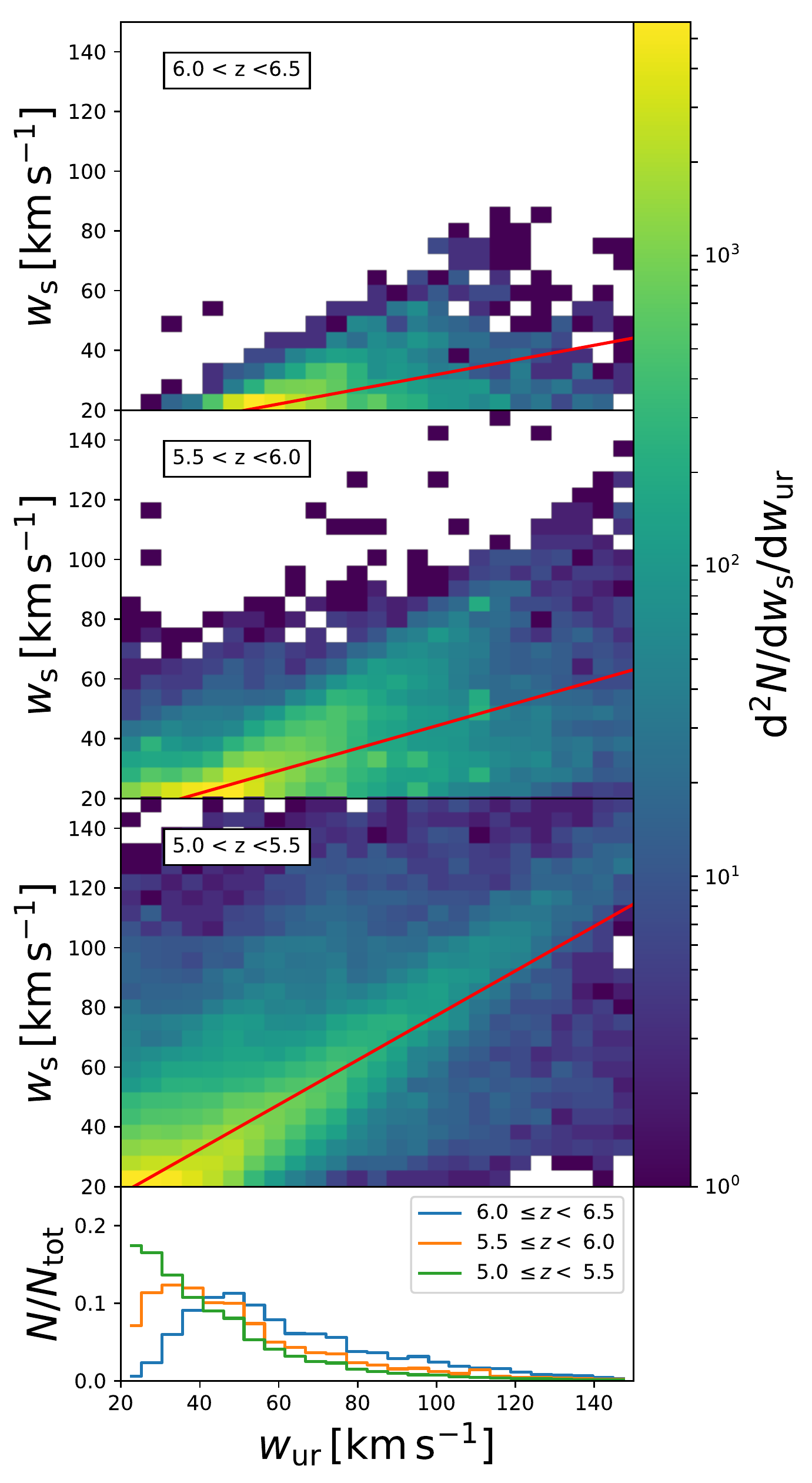}
\caption{Joint distribution of spike widths ($\wp$) and \UR lengths along the line of sight ($w_\mathrm{ur}$) in simulated spectra within three different redshift bins (top three panels). The color coding indicates the density of points in each pixel. Red lines show the best-fit linear relation between the two quantities. The bottom panel depicts the 1D normalized distribution of the sizes of \UR in the same redshift bins.}
\label{fig:void_width}
\end{figure}

Figure \ref{fig:peaks2D} shows that at any given redshift, the width of the spikes is not correlated with any of the IGM properties investigated. This naturally raises the question of what physical property sets these widths. We have shown that spikes are produced exclusively in underdense, over-ionized regions of the IGM. Hence, it appears likely that their width is simply determined by the extent of such regions along the \LOS. In order to investigate this hypothesis we identify underdense regions along the \LOS and match them with spikes as described in \S~\ref{sec:spectra}. We associate an underdensity width ($w_\mathrm{ur}$) to each spike, and investigate the connection between $\wp$ and $w_\mathrm{ur}$. The joint distribution of these two quantities is shown in Figure \ref{fig:void_width} for spikes that have an associated underdense region, color-coded with respect to the number density of spikes in any given pixel. 
For ease of comparison, we express both $\wp$ and $w_\mathrm{ur}$ in $\kms$. We
measure the former from spectra in velocity space. In the case of underdense regions, we have converted their physical length $L_\mathrm{UR}$ as
\begin{equation}
w_\mathrm{ur} =L_\mathrm{UR} \, (1+z)^{-1} \, H_0 \,  [\Omega_{\mathrm{m}} (1+z)^{3} + \Omega_{\Lambda}]^{1/2},
\end{equation}
where $L_\mathrm{UR}$ is measured in Mpc.

\begin{figure*}
\includegraphics[width=\textwidth]{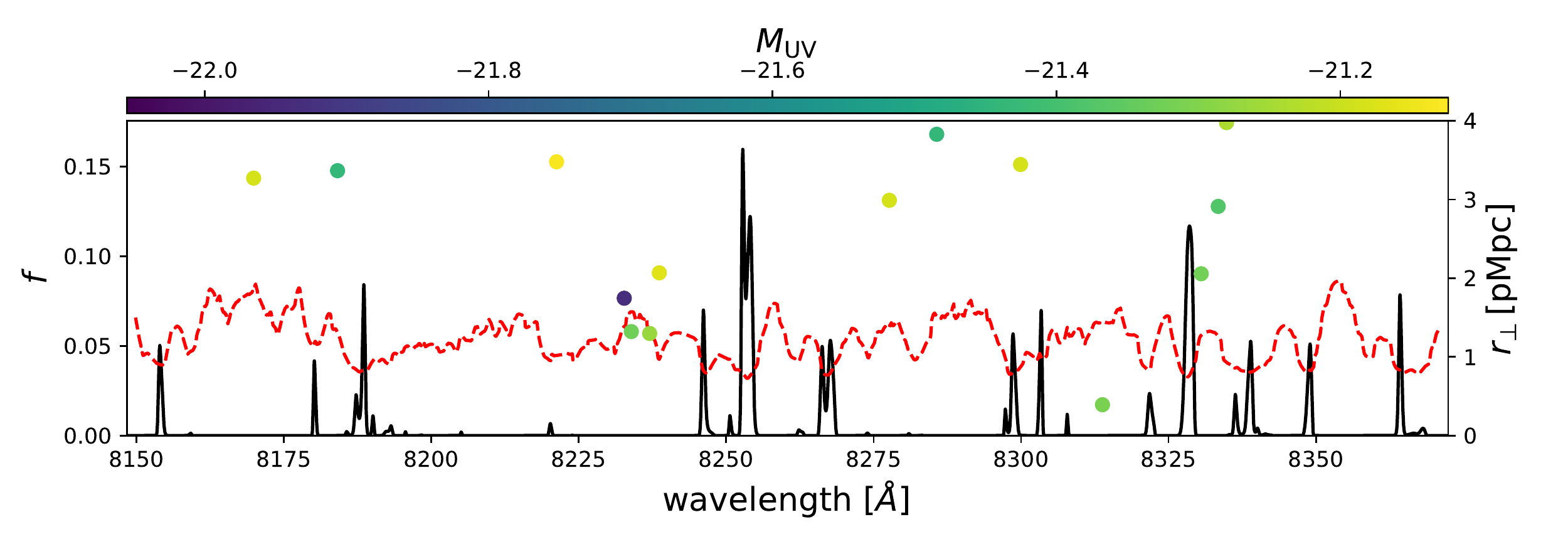}
\caption{Normalized transmitted flux along a random \LOS through our simulation box (solid line). Circles show the 
surrounding galaxies, color-coded according to their 
UV magnitude $M_\mathrm{UV}$ \enrico{\citep[computed from simulated stellar particles using the Flexible Stellar Population Synthesis library][]{FSPS2009, FSPS2010}}, 
and vertically positioned as a function of their distance from the \LOS.
The baryonic overdensity is overplotted with a dashed line on an arbitrary scale. This is a synthetic version of Figure 7 in 
\citet{Kakiichi+2018}, and for consistency we only show galaxies above their detection 
thereshold, $M_\mathrm{UV} < -21.1$.}
\label{fig:los-galaxies}
\end{figure*}

In the three redshift bins investigated, we find that the spike width is linearly correlated with the size of the \UR they originate from. The different amount of noise in the bins is fully consistent with the different number of spikes available in the spectra (see 
Figure \ref{fig:evolution}). We have confirmed this by sub-sampling the two lowest-redshift bin in such a way that they contain the same number of spikes as the highest-redshift one, resulting in all three bins having a very similar correlation strength. The slope of such correlation changes with time, however.  We show this by fitting a linear relation between $\wp$ and $w_\mathrm{ur}$ and marking the best-fitting curve as a red line in the figure. The slope of this correlation depends also on the chosen threshold for the definition of $w_\mathrm{ur}$ ($\beta$, see \S~\ref{sec:spectra}). We have investigated the effect of this parameter on the discussed correlation and show results for the value of $\beta$ that produce an approximate 1:1 relation in the lowest-redshift bin, \ie $\beta = 0.8$. We stress however that the physical interpretation of this result does not change, since varying this parameter has the only effect of changing $w_\mathrm{ur}$ at fixed $\wp$.

For completeness, we report in the bottom panel the distribution of $w_\mathrm{ur}$ \textit{associated with spikes} in the three redshift bins. Its peak shifts toward larger value with increasing redshift, a trend that can be understood as a consequence of the larger underdensities required to produce transmission spikes at earlier times (see Figure \ref{fig:peaks1D}).

\subsection{Spike-galaxy correlation}
\label{sec:spikes-galaxy}

We have shown in \S~\ref{sec:igm-prop} that transmission spikes are produced in underdense regions that are over-ionized. We interpreted this combination of physical properties as an indication that the production loci of these features are close to a UV source while still being underdense relative to the background. Similar conclusion have been obtained recently by \citet[][K18 hereafter]{Kakiichi+2018}, who
investigated observationally the concurrence of transmission spikes and neighboring galaxies by means of a spectroscopic campaign. 
Their results, although limited to a single \LOS, additionally show that the relation between spikes and nearby galaxies can be used as a tool to understand the sources of reionization and their properties. It is likely, as well as desirable, that this kind of studies will soon involve a much larger number of {\LOS}s to high-$z$ quasars. Therefore, it is of key importance to theoretically investigate this link in order to guide future observational efforts, as well as to provide theoretical predictions that can be contrasted with forthcoming data. In the following, we employ the CROC simulations to this end.

We provide a first visual impression of the concurrence of galaxies and transmission spikes in Figure\ \ref{fig:los-galaxies}, where we 
show a synthetic \Lya spectrum extracted from our simulations and the galaxies within $4\, \pMpc$ from the \LOS. The galaxy color-coding 
reflects their UV magnitude \enrico{\citep[computed from simulated stellar particles using the Flexible Stellar Population Synthesis library][]{FSPS2009, FSPS2010}}, and only objects brighter than the K18 detection threshold are shown. This figure is, effectively, a 
synthetic version of Figure 7 in K18. Also depicted in the figure is the IGM  baryonic overdensity along the \LOS, clearly 
indicating that all the observed spikes coincide with large underdense regions.
Note how the presence of a nearby bright galaxy is not a sufficient condition for the production of a spike. 
Similarly, many underdense regions are not associated with a transmission spike.

\begin{figure*}
\centering
\begin{minipage}[c]{\textwidth}
\centering
  \resizebox{\textwidth}{!}{
	\includegraphics[height=\textheight]{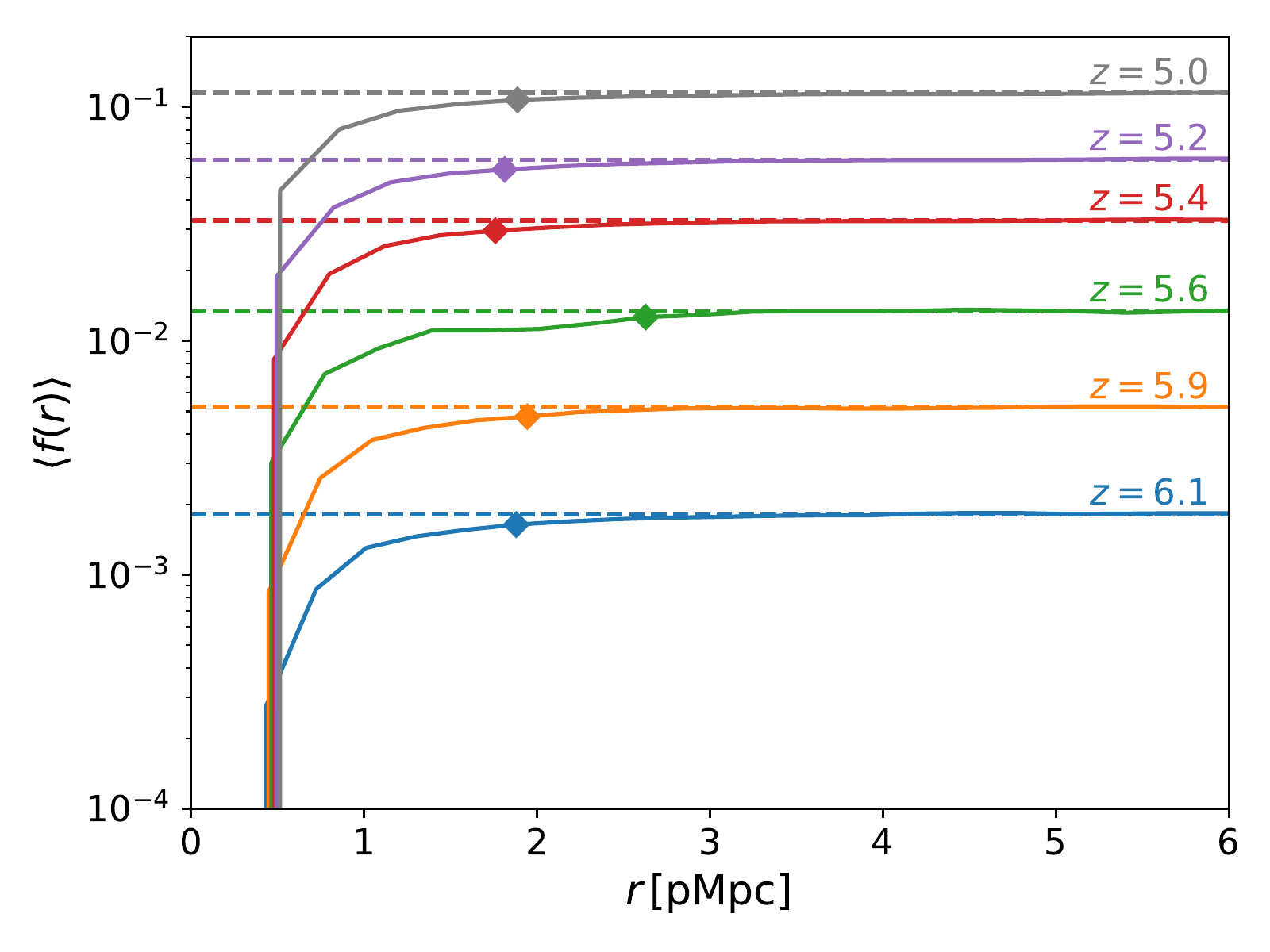}
	\includegraphics[height=\textheight]{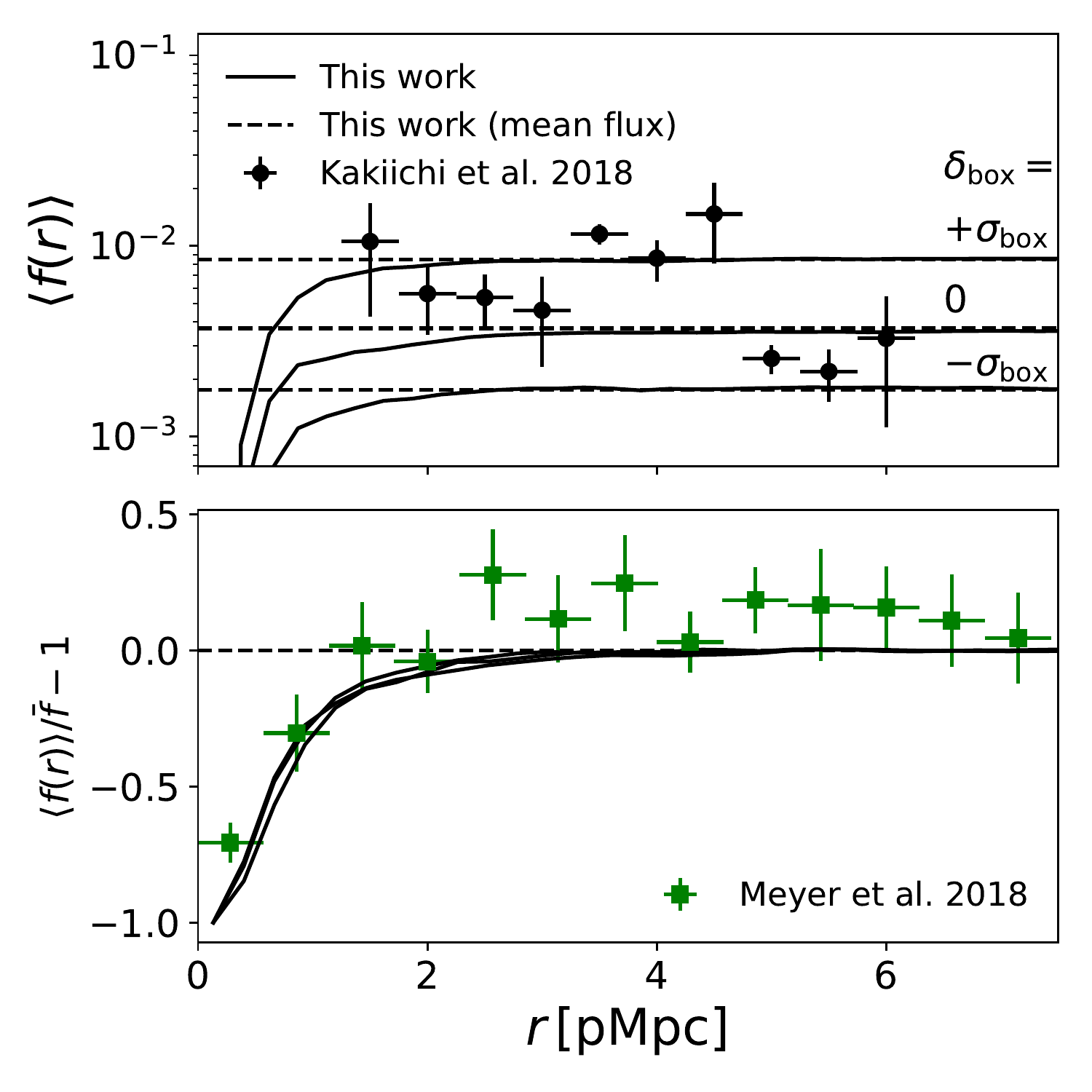}
  }
  \caption{Left: Average (normalized) transmitted flux in synthetic absorption spectra as a function of distance from the simulated galaxies. The average flux is shown for seven different redshifts (solid lines).  
  The dashed lines depict the average flux in the entire box, while diamond symbols denote where the flux reaches 95\% of the box-averaged value. Top right: Same as in the left panel but only for $z = 5.9$ and with the contribution from each simulation box plotted independently. Points with error bars show the measurements from \citet{Kakiichi+2018}. Bottom right: Difference between the average (normalized) transmitted flux at $z=5.4$ and its box-averaged value. Points with error bars show the measurements from \citet{Meyer+2019}.}
    \label{fig:avg_flux}
\end{minipage}
\end{figure*}

In order to quantify the concurrence of transmission spikes and nearby galaxies, K18 performed a correlation analysis, evaluating 
the average flux transmitted as a function of distance from nearby (spectroscopically-detected) galaxies:
\begin{equation}
\label{eq:avg_flux}
\avgf = \frac{1}{N_\mathrm{pair}(r)} \sum_{i \in \mathrm{pair}(r)} f_i  ~.
\end{equation}
Here, $N_\mathrm{pair}(r)$ is the total number of pairs in the ensemble of spectral pixel -- galaxy pairs at a given distance $r$, \ie
\begin{equation}
\mathrm{pair}(r) \equiv \{ p \in \mathcal{P}_\mathrm{LOS}, \, g \in \mathcal{G} | \, \mathrm{dist}(p,g) = r \}
\end{equation}
where $\mathcal{P}_\mathrm{LOS}$ is the ensemble of pixels in all available spectra (only one in the case of K18), $\mathcal{G}$ 
is the collection of all spectroscopically-detected galaxies, and $\mathrm{dist}(p,g)$ denotes the physical distance between a 
pixel $p \in \mathcal{P}_\mathrm{LOS}$ and a galaxy $g \in \mathcal{G}$.

We have computed the same quantity from our numerical simulations and compared it with the data and the
linear-theory predictions of K18. There are, however, few improvements with respect to their analysis, 
namely: (i) we have repeated the procedure at seven different redshifts ($z = 5.0, 5.2, 5.4, 5.6, 5.9, 6.1, 6.4$); (ii) 
we have averaged the results over $10^3$ random spectra to build each curve; and (iii) we have included all galaxies resolved in 
our simulations, which corresponds to a much lower UV detection threshold -- $M_\mathrm{UV}^\mathrm{min} \sim -16$. \enrico{For completeness, we report here that the results presented below are not affected by the latter choice, as increasing the minimum magnitude up to $M_\mathrm{UV}^\mathrm{min} \sim -21.2$ does not change the results, although it decreases their statistics. We stress, however, that $M_\mathrm{UV}^\mathrm{min}$ is just a post-processing parameters, and is not linked in any way to the minimum magnitude of galaxies producing ionizing photons \textit{in the simulation}.}

In the left panel of Figure \ref{fig:avg_flux} we show $\avgf$ extracted from our simulations. Similarly to the prediction from K18, there is $\sim 1 \, \pMpc$ region where $\avgf$ increases steadily until it saturates at exactly the average flux within the entire box, losing any information about the pixel -- galaxy distance. The drop in the average flux transmitted at $r \lesssim 1 \, \pMpc$ is consistent with being due to the effect of the overdensities that host the galaxies (see \eg Appendix B in K18). We have checked that this is indeed the case in our simulation by studying the average neutral hydrogen fraction as a function of $r$. This quantity shows a radial dependence complementary to the one of $\avgf$, \ie a sharp increase in the inner $\sim 1 \, \pMpc$ followed by a leveling off at its average value in the box.
Interestingly, the physical size of the region with suppressed flux (defined as the distance where $\avgf$ is 95 \% of the mean value) remains approximately constant with redshift (in the relatively narrow range investigated). Unlike K18, however, we do not find any region at intermediate distances where $\avgf$ is enhanced with respect to the box mean. \enrico{We will comment again on this discrepancy at the end of this paragraph, after presenting other relevant observations. In addition, we note here that computing $\avgf$ separately for galaxies of different stellar mass does not change the result. This indicates that, in our simulation, the enhanced flux coming (on average) from larger galaxies is balanced by the larger density such galaxies reside in \citep[on average, due to the tight stellar mass -- halo mass relation, \eg][]{Moster+2010}.}

The shape of the curves $\avgf$ is consistent across all the simulations, \ie for different values of $\delta_\mathrm{box}$. However, their vertical amplitude is strongly dependent on the latter. In particular, the simulation with a large scale positive (negative) density fluctuation $\delta_\mathrm{box} = +\sigma_\mathrm{box}$ ($-\sigma_\mathrm{box}$) shows a strongly suppressed (enhanced) average flux with respect to the box with $\delta_\mathrm{box} = 0$. The $\avgf$ in the latter is very close to the mean of the three simulations. This effect can be seen in the top right panel of Figure \ref{fig:avg_flux}, which we shall discuss below.
Before moving on, however, we note here that the main cause of the differences between simulations with various  DC modes is in the timing of structure formation and evolution. The offset between the different curves is in large part due to one of them ($\delta_\mathrm{box} > 0$) evolving faster than the average universe and the other  ($\delta_\mathrm{box} < 0$) evolving slower than average. We have checked that this is indeed the case by comparing $\avgf$ in different runs at the same value of $a_\mathrm{box}$ (\ie the expansion factor of the box, which is different than the universal scale factor as a consequence of the different mean density), and indeed the three boxes produce almost indistinguishable results.

In Figure \ref{fig:avg_flux} we compare our results to recent high-redshift observations by K18 (top panel) and \citet[][bottom panel]{Meyer+2019}. The measurements in these two studies are presented in different forms and cover slightly different epochs, and 
we therefore discuss them separately. To compare with the data of K18, we select from each simulation the snapshot at the epoch that is the  closest to the average redshift of the galaxies detected by K18, \ie $z \approx 5.8$. In the right panel of the figure we show the results from each simulation box separately, in order to display the effect of large-scale density fluctuations. The agreement of the simulated $\avgf$ with the observations depends strongly on the DC mode of the box. The best overall match is provided by $\delta_\mathrm{box} = 0$, in particular when
taking into account the fact that the increased observed $\avgf$ at $3 \lesssim r / \pMpc \lesssim 5$ is an artifact due to the re-sampling of a single, prominent transmission spike (K18). 
It should be noted, however, that the data of K18 are based on a single \LOS, and therefore can potentially be biased. Hence, additional observations are necessary to allow a statistically-significant statements about the ability of current reionization models to reproduce the observed average flux. 
Our results also show, however, that this task is hindered by the fact that a proper comparison between synthetic and real data requires knowledge of the large-scale overdensity of the observed field, which is mostly unavailable, especially at the high redshifts were \Lya transmission spikes are detected. Therefore, to effectively compare theoretical predictions to observed spectra, the asymptotic value of $\avgf$ needs to be factored out. This is done \eg in \citet{Meyer+2019}, where the normalized transmission is employed in place of $\avgf$. 
We note here the dependence on the DC mode can be used as a powerful tool to determine the large-scale density field from a relatively narrow patch of the sky. In fact, by comparing the flux at $r \gtrsim 2.5 \, \mathrm{pMpc}$ with its simulated value as a function of box overdensity,  we can infer the latter from observations. In particular, applying this reasoning to the observations by K18 suggests that the observed field is close to the average density of the universe on large scales.

\citet{Meyer+2019} performed a study similar to K18, but using CIV absorbers as tracers of low-mass galaxies and focusing on a slightly-lower redshift interval centered around $z \approx 5.4$. We show their results (obtained combining 25 different quasar lines of sight) in the bottom panel of Figure \ref{fig:avg_flux}. Differently from K18, they normalize the transmitted flux to the average along each line of sight. This procedure effectively erases differences due to large scale fluctuations. In fact, when we use our simulations to produce a synthetic version of such quantity, we obtain consistent results across the different simulation boxes. In particular, the outcome of our simulations is in good agreement with the observational data. Simulations, however, appear to slightly underpredict the observed values at intermediate ($2 \lesssim r / \mathrm{pMpc} \lesssim 6$) distances. \enrico{At first, this result may appear at odds with the larger-than-average ionization field experienced by the gas producing the transmission spikes (Fig.~\ref{fig:xHI_ratio}). However, the boosted ionization field around sources is balanced by an increased neutral fraction. Therefore, the presence of (a region of) boosted transmission entirely depends on the balance between these two opposing effects, which may vary with the distance from the central galaxy. }

Finally, we shall discuss here the discrepancy between our results and the linear-theory-based predictions of K18, which show a proximity zone of enhanced $\avgf$ at $r \gtrsim 2 \, \hMpc$ before approaching its asymptotic ($r \rightarrow \infty$) value at larger radii. \enrico{As detailed above, this may hint to a different balance between recombinations and photoionization at intermediate radii. In the model of K18, this boost} is a consequence of the galaxy placed at $r = 0$ that enhances the surrounding flux thanks to its radiation field. However, in deriving such prediction, K18 assumed that the galaxy located at $r=0$ is the only detected galaxy (and hence the brightest) in the range of $r$ explored. We can test if this assumption holds in our simulations. 
The faintest spectroscopically-detected galaxy in K18 has $M_\mathrm{UV}^\mathrm{thr} = -21.1$. In our simulation, galaxies above this mass threshold have, on average, $N_\mathrm{neigh} = 5$ galaxies with $M_\mathrm{UV} \geq M_\mathrm{UV}^\mathrm{thr}$ within $6 \, \pMpc$. Therefore, the region $r< 6\, \pMpc$ is affected by the proximity effect of multiple galaxies, enhancing the $\avgf$ with respect to the prediction obtained assuming a single bright source. The number $N_\mathrm{neigh}$ rapidly increases with decreasing $M_*^\mathrm{thr}$, rising to $N_\mathrm{neigh} = 21.4$ for $M_*^\mathrm{thr} = 10^9 \, \hMsol$.
Supporting this view is also the fact that the asymptotic value of $\avgf$ predicted by the model of K18 is much lower than what is obtained in our simulations, possibly indicating that this value is due to an overlap of the contribution of multiple sources.

There are, however, two possible alternative explanations.
(i) If our reionization model overestimates the galactic escape fraction, then proximity region of a galaxy extends to larger radii than in reality and, at the same time, the $\avgf$ profile becomes flatter at much shorter distances (see the bottom left panel of 
figure 8 in K18). This could indeed be the case, since the UV escape fraction at the simulation resolution ($\epsilon_\mathrm{UV}$) is a free parameter that has been calibrated to reproduce the optical depth distribution observed in the \Lya forest. The value employed $\epsilon_\mathrm{UV} = 0.15$ is in good agreement with numerical simulations of escape of ionizing radiation from molecular clouds \citep{Howard+2017,Howard+2018} at similar spatial scales (100 pc), but there exist no observational constraints on this value at present.

(ii) The second possibility is that our simulations underestimate the minimum magnitude of galaxies contributing to the ionizing photon budget. In this case, unresolved galaxies produce enough ionizing photons to increase the asymptotic $\avgf$ and suppress the proximity effect of the bright $r=0$ galaxy (see the bottom right panel of figure 8 in K18).

It is likely that a combination of all these effects is driving the difference observed in $\avgf$. Unfortunately, it is not easy to disentangle individual contributions in numerical simulations. In particular, it is highly nontrivial to predict the ionizing state of the IGM with different contribution from smaller galaxies. Similarly, changing the escape fraction will change the IGM ionization state and history in a way that is unpredictable at the level of detail required for the present study. Therefore, if the inclusion of nearby bright sources in the theoretical prediction of K18 or future observations will not solve the discrepancy, a large suite of numerical simulation with different values of relevant parameters is needed to assess which is the main reason for the difference between the analytical prediction of K18 and our results. At the moment this appears -- unfortunately -- prohibitively expensive.

\section{Summary and Conclusions}
\label{sec:conclusions}

In this work we have highlighted the information that can be extracted from the \Lya transmission regions (`spikes') embedded in the extended Gunn-Peterson troughs observed in high-redshift quasar spectra. We have used state-of-the-art CROC radiation-hydrodynamic simulations to produce synthetic spectra including the effects of gas peculiar velocity, thermal and Doppler broadening, finite spectral resolution, and instrumental noise. We employed a simple algorithm to identify spikes in the normalized transmitted flux and underdense regions along the line of sight (Figure \ref{fig:showcase}). Our results are based on three runs sharing the same initial conditions but different values of the density fluctuation on the scale of the box (the DC mode). Our results can be summarized as follows.

\begin{enumerate}
    \item The number of transmission spikes as a function of redshift evolves in the simulations from $0$ at $z \gtrsim 6$ to approximately $11 \, \mathrm{Mpc}^{-1}$ at $z = 5$. At the resolution and S/N level of current observations of the distant quasar \ULAS, this number is reduced by an order of magnitude at all redshifts. The number of spikes is larger (at all redshifts) in the run with a positive DC mode, while its smaller when the   DC mode is negative. This is a consequence of the faster (slower) evolution of the former (latter) with respect to a box having exactly the mean density of the Universe.

    \item Synthetic versions of a \ULAS-like spectrum obtained by concatenating simulation
    outputs at different redshifts show good agreement with the data in approximately 
    $15$\% of the cases.

    \item Spikes are found to originate exclusively from underdense, overly-ionized regions that become less underdense and more ionized with time. The inclusion of observational effects removes the high-density tail of the IGM underdensity distribution as the associated
    spikes have, on the average, a lower transmitted flux. 

    \item The spike width does not correlate with the IGM density at any given redshift. The spike height is negatively correlated with the gas (over)density, but with a large scatter and in a redshift-dependent fashion. The height also correlates  with the gas ionized fraction, more so at $z\approx 5.75$. Including observational effects remove the correlation between spike height and gas density, as a consequence of a preferential suppression of high-density, small-height spikes. The correlation with the gas ionised fraction is mostly unchanged (Figure \ref{fig:peaks2D_ulas}).

    \item Transmission spikes at redshift $5.0 \leq z \leq 6.5$ are produced in regions that are more ionized than expected from ionization equilibrium with the average UV background flux. Spikes are therefore formed in underdense regions of the IGM that are sufficiently close to a source of ionizing radiation.

    \item The spike width is determined by the extent of the associated underdense region
    along the line of sight. The link between width and extent changes with time (for a fixed width parameter $\beta$).

    \item The average transmission in synthetic absorption spectra as a function of distance from a bright galaxy shows a proximity zone of suppressed flux with an approximately constant radius in physical coordinates. The zone is associated with the overdense galaxy
    region. The average transmission at any given redshift depends on the value of the DC mode. Recent observations of the transmission as a function of source distance by \citet{Kakiichi+2018}
    and \citet{Meyer+2019} show some mild indications of the existence of a region with enhanced flux at intermediate radii, which is missing in our simulations. We identified two possible shortcomings of our simulations, namely the overprediction of the escape fraction or the underestimation of the minimum magnitude of galaxies that significantly contribute to the reionization of the Universe.
    
\end{enumerate}

Our analysis shows that high-redshift transmission spikes are a promising tool to push investigations of the IGM well into the tail-end of cosmic reionization. We predict such spikes to be ubiquitous in quasar sightlines, and estimate that only approximately $10$\% of all predicted spikes are detectable even with the best data currently available. Once more of these spectral features will be discovered, the way forward will ideally follow two parallel paths. On the one hand, the larger statistics will promote spikes to a powerful and complementary probe of reionization models in an era where the average \Lya optical depth is already very large \citep{Chardin+2018,Kakiichi+2018}.
On the other hand, detailed simulations that will prove able to predict the distribution and shape of the observed spikes should be used to draw a physical connection between such spectral features and the ionization and thermal state of the early IGM.

\section*{Acknowledgements}
We thank Koki Kakiichi for useful discussions on transmission spikes. 
\enrico{We also thank the anonymous referee for their comments.}
Support for this work was provided by NASA to PM through a contract to the WFIRST-EXPO Science Investigation Team (15-
WFIRST15-0004), administered by the GSFC. This research has been partially carried out within the SFB 956 
`The Conditions and Impact of Star Formation', sub-project C4, funded by the Deutsche Forschungsgemeinschaft (DFG). 
EG thanks the University of California, Santa Cruz for their warm hospitality, during which this work was conceived and initiated.
Fermilab is operated by Fermi Research Alliance, LLC, under Contract No. DE-AC02-07CH11359 with the United States Department of Energy. This work was partly supported by a NASA ATP grant NNX17AK65G, and used resources of the Argonne Leadership Computing Facility, which is a DOE Office of Science User Facility supported under Contract DE-AC02-06CH11357. An award of computer time was provided by the Innovative and Novel Computational Impact on Theory and Experiment (INCITE) program. This research is also part of the Blue Waters sustained-petascale computing project, which is supported by the National Science Foundation (awards OCI-0725070 and ACI-1238993) and the state of Illinois. Blue Waters is a joint effort of the University of Illinois at Urbana-Champaign and its National Center for Supercomputing Applications. We made extensive use of the NASA Astrophysics Data System and {\tt arXiv.org} preprint server, and are thankful to the community developing and maintaining the software 
packages Matplotlib \citep{matplotlib}, NumPy \citep{numpy}, SciPy \citep{scipy}.

\bibliographystyle{aasjournal}
\bibliography{spikes}

\appendix
\section{Simulating \ULAS}
\label{sec:AppendixA}
Many of the results discussed in this work are instrument-agnostic, based on synthetic spectra that have a very high spectral resolution and no noise. This has allowed us to produce an unbiased view of the physical mechanisms at the origin of the \Lya transmission spikes and, at the same time, to 
produce products that can easily be translated in predictions for any given combination of spectral resolution and noise level. Here, we assess
the impact of realistic observational effects using the observations of \ULAS \citep{Barnett+2017} as a template (see \S~\ref{sec:number_evol}). In 
particular, we present here the `observer version' of Figures \ref{fig:peaks1D} and \ref{fig:peaks2D}. 

\begin{figure*}
\includegraphics[width=\textwidth]{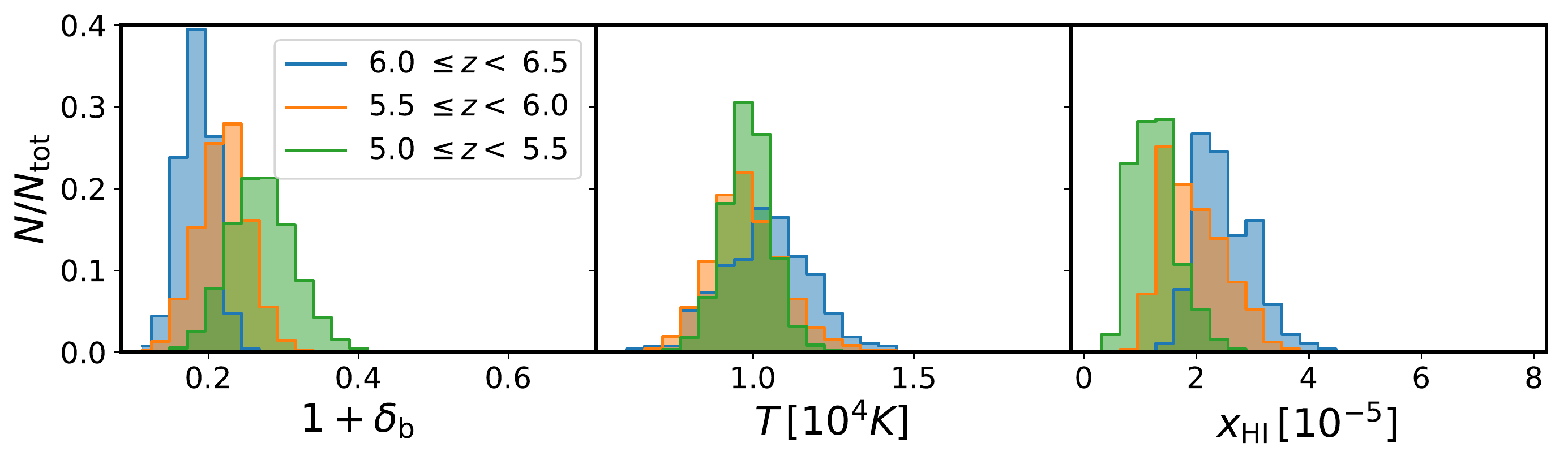}
\caption{Same as Figure \ref{fig:peaks1D}, but for \ULAS-like spectral resolution and noise.}
\label{fig:peaks1D_ulas}
\end{figure*}

{Figure \ref{fig:peaks1D_ulas} shows the properties of the intergalactic gas responsible for the  transmission spikes. A comparison with its idealized counterpart (Figure \ref{fig:peaks1D}) reveals that observational effects preferentially erases spikes originating from the less-underdense regions. This is a consequence of finite spectral resolution smearing out narrow spikes and reducing their height. The latter is smaller in less-underdense regions (see Figure \ref{fig:peaks2D}), which are therefore preferentially suppressed once a noise threshold is imposed.}

\begin{figure*}
\includegraphics[width=\textwidth]{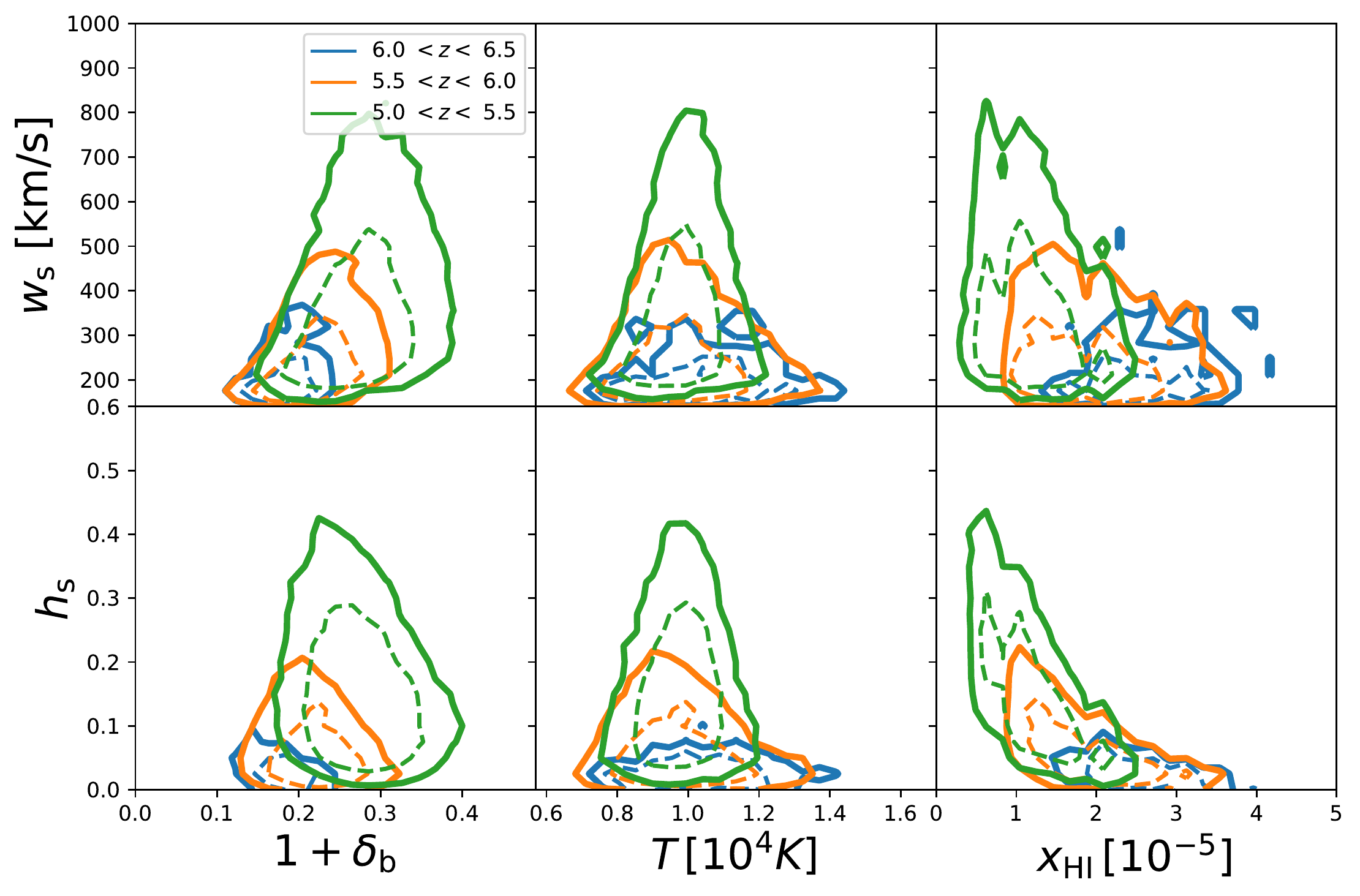}
\caption{Same as Figure \ref{fig:peaks2D}, but for \ULAS-like spectral resolution and noise.}
\label{fig:peaks2D_ulas}
\end{figure*}

{Figure \ref{fig:peaks2D_ulas} is the `observer version' of Figure \ref{fig:peaks2D} (notice that the axes have different ranges, for the sake of visual clarity). The inclusion of realistic observational conditions has a two-fold impact. The first is the shrinking of the allowed properties for the gas producing detectable spikes. The second is due to finite spectral resolution broadening the spikes and stretching the range of spike widths by more than a factor of $5$. This effects also induce a cut-off in the minimum $\wp$ at approximately $200 \, \kms$, which is larger than the spectral resolution ($\Delta v = 10 \, \kms$). Resolution suppresses the flux of narrow spikes as they are smeared over a larger spectral range, so that only broad spikes survive the noise threshold selection. An exception to this condition
is represented by spikes that are very narrow but also very high, so that even after being smeared out they are still detectable above the noise. Such spikes are rare and therefore are not contained in the 95 per cent contour shown in the figure. We note here that observational effects do not change significantly the results of Figure \ref{fig:xHI_ratio}, although they reduce the overall number of detectable spikes as already discussed.}

\end{document}